\documentclass[english,12pt]{article}
\usepackage{array}
\usepackage{graphicx}
\usepackage{amssymb}
\usepackage[english]{babel}
\usepackage{amsmath}
\usepackage{multirow}
\usepackage{prettyref}
\usepackage{babel}
\usepackage{units}
\usepackage[latin1]{inputenc}
\usepackage{amsfonts}
\usepackage{amssymb}
\usepackage{babel}
\usepackage{color}
\usepackage{cite}
\def\@fmsl@sh#1#2#3{\m@th\ooalign{$\hfil#1\mkern#2/\hfil$\crcr$#1#3$}}
 \def\eq#1\en{\begin{equation}#1\end{equation}}
\def\s[#1,#2]{[#1\stackrel{\star}{,}#2]}
\def\sx[#1,#2]{[#1\stackrel{\star_{x}}{,}#2]}

\newcommand{\nc}{\newcommand}
\nc{\beq}{\begin{equation}}
\nc{\eeq}{\end{equation}}
\nc{\beqa}{\begin{eqnarray}}
\nc{\eeqa}{\end{eqnarray}}

\def\bc{\begin{center}}
\def\ec{\end{center}}

\def\to{\rightarrow}

\def\gsim{\mathrel{\mathpalette\atversim>}}

\def\bc{\begin{center}}
\def\ec{\end{center}}

\def\gsim{\mathrel{\rlap{\lower4pt\hbox{\hskip1pt$\sim$}}

    \raise1pt\hbox{$>$}}}       

\def\gsim{\mathrel{\rlap{\lower4pt\hbox{\hskip1pt$\sim$}}
    \raise1pt\hbox{$>$}}}       

\begin{document}
\makeatletter
\def\fmslash{\@ifnextchar[{\fmsl@sh}{\fmsl@sh[0mu]}}
\def\fmsl@sh[#1]#2{%
  \mathchoice
    {\@fmsl@sh\displaystyle{#1}{#2}}%
    {\@fmsl@sh\textstyle{#1}{#2}}%
    {\@fmsl@sh\scriptstyle{#1}{#2}}%
    {\@fmsl@sh\scriptscriptstyle{#1}{#2}}}
\def\@fmsl@sh#1#2#3{\m@th\ooalign{$\hfil#1\mkern#2/\hfil$\crcr$#1#3$}}
\makeatother

\thispagestyle{empty}
\begin{titlepage}
\boldmath
\begin{center}
  \Large {\bf  On Searches for Gravitational Dark Matter with Quantum Sensors}
    \end{center}
\unboldmath
\vspace{0.2cm}
\begin{center}
{  {\large Xavier Calmet}\footnote{x.calmet@sussex.ac.uk}}
 \end{center}
\begin{center}
{\sl Department of Physics and Astronomy, 
University of Sussex, Brighton, BN1 9QH, United Kingdom
}\end{center}
\vspace{5cm}
\begin{abstract}
\noindent
The possibility of searching for dark matter with quantum sensors has recently received a lot of attention. In this short paper, we discuss the possibility of searching for gravitational dark matter with quantum sensors and identify a very narrow window of opportunity for future quantum sensors with improved sensitivity. Gravitational dark matter candidates with masses in the range $[10^{-3}, 1] \, \text{eV}$ could lead to an effective time variation of the proton mass that could be measured with, e.g., future atomic clocks.
\end{abstract}  
\vspace{5cm}
\end{titlepage}



\newpage

Understanding the nature of dark matter remains one of the most profound challenges in modern physics. We know remarkably little about dark matter besides the fact that it constitutes about 85$\%$ of the matter density in our universe. Furthermore, according to sophisticate numerical simulations, it must be non-relativistic, or cold, to account for the formation of galaxy. For a recent reviews, see e.g. \cite{Tanabashi:2018oca} and \cite{Garrett:2010hd}. For galaxies to form, dark matter must be heavier than $10^{-22}$ eV \cite{Hu:2000ke}.

Despite convincing astrophysical and cosmological observations that have accumulated for decades confirming the existence of a non-luminous form of matter, the nature of dark matter particles remains a complete mystery as one thing is clear, there is no viable dark matter candidate within the standard model of particle physics. The mass range for dark matter particles and the strength of their coupling constants to the particles of the standard model remains widely model dependent and there are no known generic features. 

While searches have essentially focussed on collider searches or experiments involving the recoil of nuclei when being hit by a dark matter particle traveling through space, it has been recently realized that quantum sensors could play an important role in the search for dark matter candidates \cite{Arvanitaki:2014faa,Hees:2016gop,Arvanitaki:2015iga,Graham:2015ifn,Graham:2012sy,Arvanitaki:2016fyj,Stadnik:2014ala,Stadnik:2014tta,Roberts:2015lma,Stadnik:2018sas}.  In particular atomic clocks could probe an interesting mass range for very light dark matter.

In this paper, we will focus on particles that couple to the energy-momentum tensor of the Standard Model which appear in all models on quantum gravity. We call this class of dark matter candidates gravitational dark matter. This class of models is actually rather  broad, it incorporates dark matter candidates motivated by modified gravity see e.g.\cite{Cembranos:2008gj,Calmet:2017voc} and quantum general gravity \cite{Calmet:2018uub}. Extensions of general relativity and quantum gravity have rather generically new scalar degrees of freedom that couple to the trace of the energy momentum tensor of the standard model (see Appendix A):
\begin{align}\label{Eq1}
S=\int d^4 x \left[  \frac{1}{2} \partial_\mu \sigma  \partial^\mu \sigma  - \frac{m_\sigma^2}{2} \sigma^2 - \sqrt{\frac{8 \pi G_N}{3}}  \sigma \eta_{\mu\nu}T^{\mu\nu} 
  \right ],
\end{align}
where $\sigma$ is a massive spin-0 field and $T^{\mu\nu}$ is the energy-momentum tensor of the standard model. In the specific case of quantum general relativity, it was shown in \cite{Calmet:2018uub} that this field is viable dark matter candidates if its masses is in the interval $[1\times 10^{-12},0.16]\ \mbox{GeV}$. 

We are interested in the couplings of $\sigma$  to leptons, quarks, gluons and the photon. The energy momentum tensor for the photon is given by
\begin{eqnarray}
T^{\mu\nu}=F^{\mu\alpha}F^{\nu}_{\ \alpha}-\frac{1}{4} \eta^{\mu\nu} F_{\alpha\beta}F^{\alpha\beta}
\end{eqnarray}
and by 
\begin{eqnarray}
T^{\mu\nu}=i  \bar \psi \gamma^\mu \partial^\nu \psi - \eta^{\mu\nu} \bar \psi (i \gamma^\alpha \partial_\alpha-m) \psi 
\end{eqnarray}
for fermions. We see that the scalar field $\sigma$ will not couple to the photon at tree level because the energy momentum tensor is traceless. The same applied to the gluons. However, such a coupling will be induced at one-loop \cite{Giudice:2000av}:
 \begin{eqnarray} 
\sqrt{\frac{8 \pi G_N}{3}} \left (b_2+b_Y-F_1(\tau_W)+\sum_f N_{C,f} \left(\frac{Q_f}{3}\right)^2 F_{1/2}(\tau_f)\right)
\sigma \frac{\alpha}{8 \pi} F_{\alpha\beta}F^{\alpha\beta},
\end{eqnarray}
where the sum runs over all fermions that couple to $\sigma$, $\alpha=1/137$ $b_2 = 19/6$, $b_Y = 41/6$, $N_{C,f}$ is equal to 3 for quarks and 1 for leptons and finally $Q_f$ is the QED-charge of the fermion. The one-loop form factors $F_1(\tau_W)$ and $F_{1/2}(\tau_f)$, where $\tau_W=4 m_W^2/m_\sigma^2$ ($m_W$ is the mass of the W-bosons) and $\tau_f=4 m_f^2/m_\sigma^2$ ($m_f$ is the mass of the fermion in the loop) have the following limit when $m_f$ or $m_W$  is much larger than $m_\sigma$:
 \begin{eqnarray} 
 F_{1/2}(\infty) \to -\frac{4}{3}
\end{eqnarray}
and
 \begin{eqnarray} 
 F_{W}(\infty) \to 7.
\end{eqnarray}
We can see that if $\sigma$ is much lighter than all the fermions of the standard model and the $W$-bosons, $\sigma$ decouples from the electromagnetic field at 1-loop as well:
$19/6 - 41/6 - 7 - (-4/3) (3 \times 3 \times (2/3)^2 + 3 \times 3  \times (-1/3)^2 + 3  \times (-1)^2)=0$.

A coupling to the gluons will also be induced at one loop as well:
 \begin{eqnarray} 
\sqrt{\frac{8 \pi G_N}{3}} \left (b_3- \frac{1}{2} \sum_q  F_{1/2}(\tau_q)\right)
\sigma \frac{\alpha_S}{8 \pi} G^a_{\alpha\beta}G^{a\alpha\beta},
\end{eqnarray}
where the sum runs over all quarks $q$ which are coupling to $\sigma$, $\alpha_S$ is the strong coupling constant and $b_3 = 7$ is the QCD $\beta$-function coefficient in the standard model. We see that if all quarks are heavier than the $\sigma$ field $\left (b_3- \frac{1}{2} \sum_q  F_{1/2}(\infty)\right)\to 11$.

We can thus consider the following effective Lagrangian
\begin{eqnarray} 
 L&=&  d_\gamma \frac{\sqrt{4 \pi G_N}}{4} \sigma F_{\alpha\beta}F^{\alpha\beta}+d_g \frac{\sqrt{4 \pi G_N}}{4} \sigma G^a _{\alpha\beta}G^{a \alpha\beta} 
 \\ \nonumber &&
 - d_e\sqrt{4 \pi G_N} \sigma m_e \bar e e- d_q\sqrt{4 \pi G_N} \sigma  m_q \bar q q \
\end{eqnarray}
to describe our generic gravitational dark matter particle $\sigma$ to the photon, gluons, electrons and quarks. In our model, we have 
\begin{eqnarray} 
d_e&=&d_q=4 \sqrt{\frac{2}{3}}\approx 3.3 \\
d_\gamma&=&3 \sqrt{\frac{2}{3}}  \left (b_2+b_Y-F_1(\tau_W)+\sum_f N_{C,f} \left(\frac{Q_f}{3}\right)^2 F_{1/2}(\tau_f)\right) \\
d_g&=& 3 \sqrt{\frac{2}{3}} \left (b_3- \frac{1}{2} \sum_q  F_{1/2}(\tau_q)\right).
\end{eqnarray}

Although we started from a gravitational theory which has only one free parameter namely Newton's constant, we do not necessarily end up with a universal coupling of the dark matter candidate to matter.  There is thus a range of values for $d_i$ that are not necessarily equal to one despite starting from a gravitational interaction which is expected to be the weakest force in nature. We see that for fifth force types of interactions, the coupling to matter can be weaker than the gravitational interaction. However, we can also see that the coupling to leptons and quarks are universal. The bound from the E\"ot-Wash experiment \cite{Kapner:2006si,Hoyle:2004cw,Adelberger:2006dh} thus applies to the mass of any dark matter candidate of gravitational origin that couples to $T$ which must be heavier than  $1\times 10^{-3}  \mbox{eV}$. We now discuss the possibility to discover such dark matter candidates with quantum sensors such as atomic clocks.

If the scalar field is the main component of dark matter it is easy to estimate the number of particles per reduced de Broglie volume. Given the local dark matter energy density $\rho_{DM}=0.3$ GeV/cm$^3$, one can show that if the scalar field is lighter than 1~eV then there is a large number of particles per reduced de Broglie volume. This implies that the scalar field behaves as a highly classical state and it can be approximated by a non-relativistic plane wave 
\begin{align}
\phi\left(t,\vec{x}\right) = \phi_0 \cos\left[m_\phi (t -  \vec{v}\cdot\vec{x}) + \beta\right]+\mathcal{O}\left(|\vec{v}|^2\right), 
\end{align}
where the amplitude is given by $\phi_0 \simeq \sqrt{2\rho_\text{DM}}/{m_\phi}$. It is determined by the local dark matter energy density. The oscillations of the  scalar fields lead to an effective time dependence of the  fine-structure constant
\begin{align}
\alpha(t,\vec{x}) &= \alpha \left[1 + d_\gamma \sqrt{4\pi G_N} \phi(t,\vec{x})\right]
\end{align} 
and of the QCD coupling constant (and hence the QCD scale)
\begin{align}
\alpha_S(t,\vec{x}) &= \alpha_S \left[1 + d_{g} \sqrt{4\pi G_N} \phi(t,\vec{x})\right].
\end{align} 
The same applies to the lepton masses and quark masses
\begin{align}
m_e(t,\vec{x}) &=m_e \left[1 + d_e \sqrt{4\pi G_N} \phi(t,\vec{x})\right],\\
m_q(t,\vec{x}) &= m_q \left[1 + d_q \sqrt{4\pi G_N} \phi(t,\vec{x})\right].
\end{align} 
As the scalar fields couple to the quarks and to the gluons as well, a similar effect would be seen for the proton mass $m_p$. It is essentially fixed by the the QCD scale and thus gluon dynamics. The QCD scale depends on $\alpha_S$ and is given by:
\begin{align}
\Lambda_{QCD}=\mu \sqrt{ \exp \left ( \frac{4 \pi}{c_{G} \alpha_S}\right)},
\end{align} 
which varies in time if $\alpha_S$ is time dependent, we find
\begin{align}
\Lambda_{QCD}(t,\vec{x})=\Lambda_{QCD} \left (1-\frac{2 \pi}{c_{G} \alpha_S} d_{g} \sqrt{4\pi G_N} \phi(t,\vec{x})\right).
\end{align} 
The time dependence of the proton mass is thus given by
\begin{align} \label{protonmass}
m_p(t,\vec{x}) &= m_p \left (1-\frac{2 \pi}{c_{G} \alpha_S} d_{g} \sqrt{4\pi G_N} \phi(t,\vec{x})\right),
\end{align} 
where $c_{G}=b_3$.

As we have seen the mass of the dark matter field needs to be below 1~eV to be classical enough to generate the time variation effect that would be detedectable with quantum sensors sensitive to a change in $\alpha$, while the fields must have masses larger than $1\times 10^{-3} \ \mbox{eV}$ to avoid bounds form the E\"ot-Wash experiment. We thus find that their masses are within the interval $[10^{-3}, 1] \, \text{eV}$.  which is a very narrow window. Furthermore, for such light dark matter fields, fermions will decouple from the loop leading to interactions with photons and gluons. We find $d_\gamma=0$ and $d_g=11\sqrt{6}$. In other words, extremely light scalar fields that couple to the trace of the energy momentum tensor do not lead to a time variation of $\alpha_{QED}$, however they will lead to a time variation of the fermion masses, QCD scale and hence of the proton mass. However, using the sensitivity of optical or microwave clocks presented in \cite{Arvanitaki:2014faa}, it is easy to see that current quantum sensors cannot look for dark matter candidate of gravitational origin. They would also be out of reach of future \cite{Arvanitaki:2016fyj} atomic gravitational wave detectors. Here we have focussed on scalar field dark matter, there are also spin-2 candidates in gravitational theories, see e.g. \cite{Calmet:2018uub,Marzola:2017lbt}. It is however easy to show that the same conclusion applies to these higher spin bosons.

A similar class of models are string dilaton ones \cite{Callan:1985ia,Callan:1986jb,Kaplunovsky:1987rp,Damour:1994zq}. These model have a coupling between the dilaton $\phi$ and the kinetic term of the photon:  $\sqrt{G_N} \phi  F_{\mu\nu} F^{\mu\nu}$ with similar couplings to the gluons and fermion masses. As there is no further  suppression factor to make these interactions weaker than gravity, the dilaton needs to be heavier than $10^{-3}$ eV and is therefore essentially irrelevant for searches using quantum sensors.

We thus see that well motivated models for very light gravitational dark matter candidates fail to be relevant for searches with currently available quantum sensors. However, there is a narrow window $[10^{-3}, 1] \, \text{eV}$ could be probed in the future if quantum sensors, such as e.g. atomic clocks, can improve their sensitivity.

{\it Acknowledgments:}
This work supported in part  by the Science and Technology Facilities Council (grant number ST/P000819/1). 


\section*{Appendix A}
In this appendix we show that Eq. (\ref{Eq1}) follows from the mapping of the simplest possible modified gravity action
\begin{eqnarray}
S_{grav} &=& \int d^4x \, \sqrt{-g} \left[  \frac{1}{2}  M_P^2  R+c_1R^2 + \mathcal{L}_{SM}   \right],
\end{eqnarray}
to the Einstein frame. There is thus no freedom in the choice of the interaction between the scalar field and the energy momentum tensor of the standard model.

This is an example of a $f(R)$ theory with $f(R)=R+2 c_1 R^2/M_P^2$. It is well known that after a Legendre transformation followed by a conformal rescaling $\tilde{g}_{\mu\nu} = f'(R) g_{\mu\nu}$, the $f(R)$ theory can be put in the form \cite{DeFelice:2010aj}
\begin{eqnarray}
S &=& \int d^4x \sqrt{-\tilde g} \left( \frac{1}{2}  M_P^2 \tilde{R} - \frac{1}{2}\tilde{g}^{\mu\nu}\partial_\mu\sigma\partial_\nu\sigma - V(\phi)\right)\nonumber\\
&&+ \int d^4x \sqrt{-\tilde g} F^{-2}(\sigma)\mathcal{L}_M(F^{-1}(\sigma)\tilde g_{\mu\nu},\psi_M),
\label{eq:conf}
\end{eqnarray}
where
\begin{eqnarray}
\sigma\equiv \sqrt{\frac{3}{2}} M_P \log F,\\
F(\sigma)\equiv f'(R(\sigma)).
\end{eqnarray}
Hence all the matter fields acquires a universal coupling to a new scalar field $\sigma$ through the factor $F^{-1}(\sigma)$. Gauge bosons are exceptions since their Lagrangians are invariant under the metric rescaling, couplings can be generated at the 1-loop level. Linearizing this equation, we recover Eq.(\ref{Eq1}). 
\begin{align}
S=\int d^4 x \left[  \frac{1}{2} \partial_\mu \sigma  \partial^\mu \sigma  - \frac{m_\sigma^2}{2} \sigma^2 - \sqrt{\frac{8 \pi G_N}{3}}  \sigma \eta_{\mu\nu}T^{\mu\nu} 
  \right ].
\end{align}

\bigskip{}

\end{document}